\documentclass[
  ,draft            
  ]
  {aipproc}

\layoutstyle{8x11double}

\begin{document}

\title{Three Family Models from the Heterotic String}


\author{Stuart Raby}{
  address={Department of Physics, The Ohio State University, 191 W. Woodruff Ave., Columbus, OH 43210, USA}
}

\begin{abstract}
 In this talk I outline work done in collaboration with R.J. Zhang and T. Kobayashi.   We show how to construct
 the equivalent of three family orbifold GUTs in five dimensions from the heterotic string.   I focus on one
 particular model with E(6) gauge symmetry in 5D, the third family and Higgs doublet coming from the 5D bulk and
 the first two families living on 4D SO(10) branes.   Note the E(6) gauge symmetry is broken to Pati-Salam in 4D
 which subsequently breaks to the Standard Model gauge symmetry via the Higgs mechanism.   The model has two
 flaws, one fatal and one perhaps only unaesthetic.   The model has a small set of vector-like exotics with
 fractional electromagnetic charge.   Unfortunately not all of these states obtain mass at the compactification
 scale.  This flaw is fatal.   The second problem is R parity violating interactions.   These problems may be
 avoidable in alternate orbifold compactification schemes.   It is these problems which we discuss in this talk.
\end{abstract}

\maketitle


\section{Phenomenological guidelines}
We used the following guidelines when searching for ``realistic" string
models~\cite{Kobayashi:2004ud,Kobayashi:2004ya}.   We want to:
\begin{enumerate}
\item Preserve gauge coupling unification; \item  Low energy SUSY as solution to the gauge hierarchy problem,
i.e. why is $M_Z << M_G$; \item Put quarks and leptons in {\bf 16} of SO(10); \item Put Higgs in {\bf 10}, thus
quarks and leptons are distinguished from Higgs by their SO(10) quantum numbers; \item Preserve GUT relations for
3rd family Yukawa couplings; \item  Use the fact that GUTs accommodate a ``Natural" See-Saw scale ${\cal
O}(M_G)$, and  \item Use intuition derived from Orbifold GUT constructions.
\end{enumerate}
It is the last guideline which is novel and characterizes our approach.

\subsection{$E_8 \times E_8$ 10D heterotic string compactified on $Z_3 \times Z_2$ 6D orbifold}

\begin{figure}
  \includegraphics[height=.08\textheight]{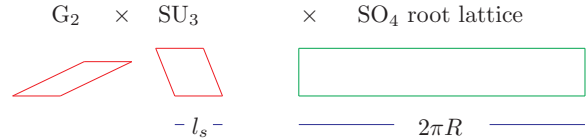}
  \caption{6D torus defined by the $G_2 \times SU_3 \times SO_4$ root lattices.  Note, one cycle has a length,
  $2 \pi R$, which is much larger than all other cycles with length given by the string length, $l_s$.}
  \label{f:torus}
\end{figure}

\begin{figure}
  \includegraphics[height=.1\textheight]{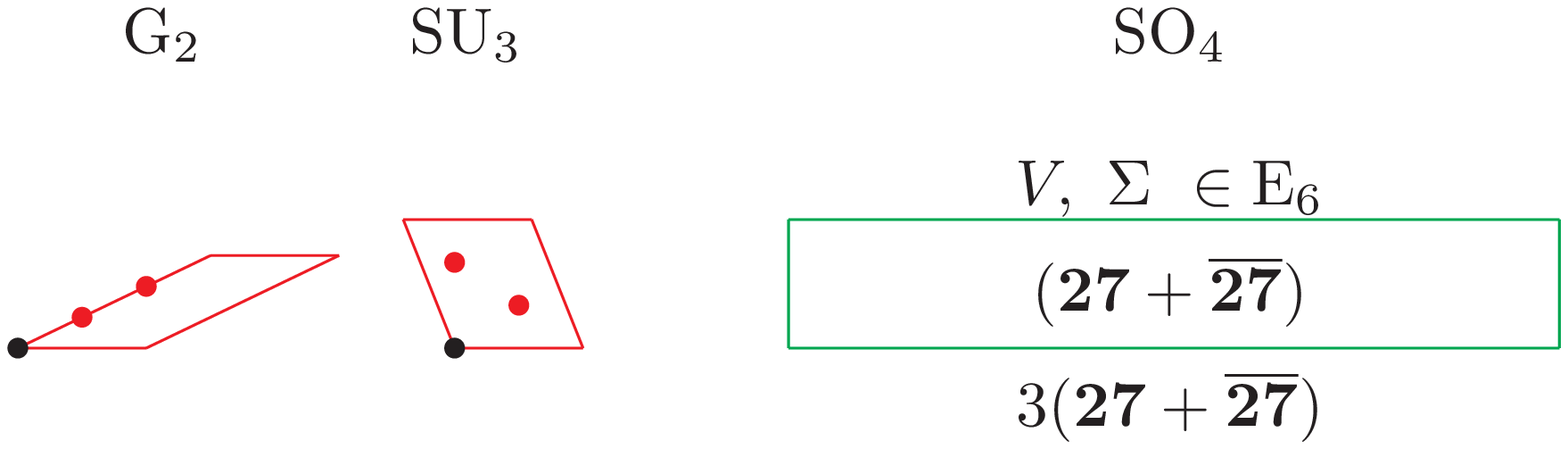}
  \caption{$(T^2)^3/Z_3$ fixed points.  Note the $Z_3$ twist does not act on the $SO_4$ torus.}
  \label{f:z3}
\end{figure}

We start with the $E_8 \times E_8$ 10D heterotic string compactified on a 6D torus $(T^2)^3$ as shown if Fig.
\ref{f:torus}.  We then mod out by the point group $Z_6 \equiv Z_3 \times Z_2$ and also add Wilson
lines~\cite{het}.
We consider this in two steps.  Let us first consider the orbifold $(T^2)^3/Z_3$ plus the Wilson line $W_3$ in
the $SU_3$ torus.  The $Z_3$ twist does not act on the $SO_4$ torus, see Fig. \ref{f:z3}.  As a consequence of
embedding the $Z_3$ twist as a shift in the $E_8 \times E_8$ group lattice and taking into account the $W_3$
Wilson line, the first $E_8$ is broken to $E_6$.  In addition we find the massless states  $\Sigma \in {\bf 78}$,
and ${\bf 27} + \overline{\bf 27}$ in the 6D untwisted sector.  These form a complete N=2 gauge multiplet and
{\bf 27} dimensional hypermultiplet. In the twisted sector we find the three massless {\bf 27} dimensional
hypermultiplets, $3 ({\bf 27} + \overline{\bf 27})$, sitting at the 3 $G_2$ fixed points and at the origin in the
$SU_3$ torus. Nevertheless, the massless states in this sector can all be viewed as bulk states moving around in
a large 5D space-time.

\begin{figure}
  \includegraphics[height=.06\textheight]{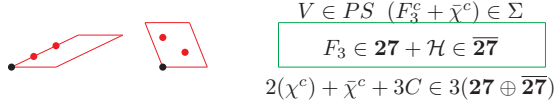}
  \caption{$(T^2)^3/Z_3$ fixed points.  The massless states in the untwisted sector and at the $G_2, \ SU_3$ fixed points
  after the $Z_2$ twist and $W_2$ Wilson line.}
  \label{f:z2}
\end{figure}

\begin{figure}
  \includegraphics[height=.2\textheight]{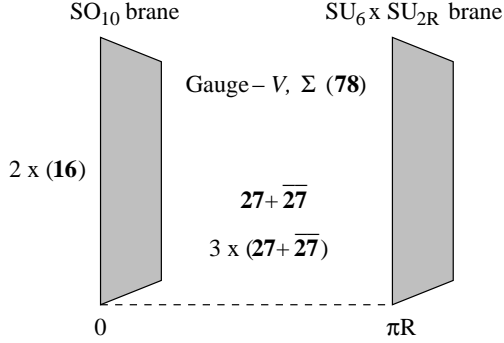}
  \caption{$E_6$ orbifold GUT in 5D with end of world branes.  The gauge symmetry at $y=0$ ($y = \pi R$) is $SO_{10}$
  ($SU_6 \times SU_{2R}$).}
  \label{f:orb}
\end{figure}

Now consider the $Z_2$ twist and the Wilson line $W_2$ along the long cycle in the $SO_4$ torus.   The action of
the $Z_2$ twist breaks the gauge group to SO(10), while $W_2$ breaks SO(10) further to Pati-Salam $SU(4)_C \times
SU(2)_L \times SU(2)_R$.   The massless spectrum from the bulk is given in Fig. \ref{f:z2}.   Out of the
untwisted sector we identify the third family of quarks and leptons with $F_3 = (4, 2, 1)$ and $F^c_3 = (\bar 4,
1, 2)$ and the Higgs doublets ${\cal H} = (1, 2, 2)$.  The color triplet Higgs have been projected out by the
Wilson line.
To summarize, the bulk physics is equivalent to an $E_6$ orbifold GUT in 5D with four N=2 hypermultiplets in the
${\bf 27}$ dimensional representation.  We shall see that the $Z_2$ fixed points in the $SO_4$ torus may be
interpreted as the end of world branes at $y = 0, \pi R$ of the orbifold GUT (see Fig. \ref{f:orb}).

\begin{figure}
  \includegraphics[height=.1\textheight]{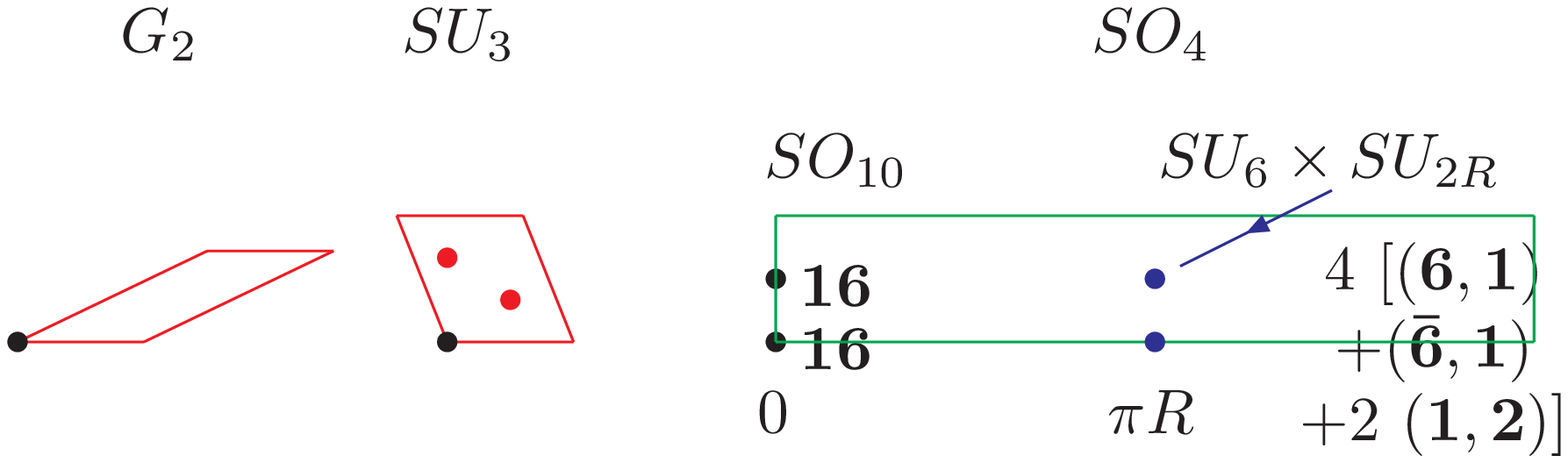}
  \caption{In the $Z_2$ twisted sector the left two fixed points are $SO_{10}$ invariant, while the two
  right fixed points are $SU_6 \times SU_{2R}$ invariant.  The overlap is Pati-Salam.  We find two families, one
  on each of the $SO_{10}$ fixed points.   On the other fixed points we find a small set of vector-like exotics. }
  \label{f:2fam}
\end{figure}

Consider the $Z_2$ fixed points.   We have four fixed points, separated into an $SO_{10}$ and $SU_6 \times
SU_{2R}$ invariant pair by the $W_2$ Wilson line (see Fig. \ref{f:2fam}).   We find two complete families, one on
each of the $SO_{10}$ fixed points and a small set of vector-like exotics (with fractional electric charge) on
the other fixed points. Since $W_2$ is in the direction orthogonal to the two families, we find a non-trivial
$D_4$ family symmetry. This will affect a possible hierarchy of fermion masses.   We will discuss the family
symmetry and the exotics in more detail next.
The discrete group $D_4$ is a non-abelian discrete subgroup of $SU_2$ of order 8.  It is generated by the set of
$2 \times 2$ Pauli matrices
\begin{equation}
D_4 =  \{ \pm 1, \pm \sigma_1, \pm \sigma_3,  \mp i \sigma_2 \}.
\end{equation}
In our case, the action of the transformation  $\sigma_1 = \left( \begin{array}{cc} 0 & 1 \\ 1 & 0 \end{array}
\right)$ takes $F_1 \leftrightarrow F_2$,  while the action of $\sigma_3 = \left( \begin{array}{cc} 1 &0 \\ 0 &
-1
\end{array} \right)$ takes  $F_2 \rightarrow - F_2$.   These are symmetries of the string.  The first is an
unbroken part of the translation group in the direction orthogonal to $W_2$ in the $SO_4$ torus and the latter is
a stringy selection rule resulting from $Z_2$ space group invariance.   Under $D_4$ the three families of quarks
and leptons transform as a doublet, ($F_1, \ F_2$), and a singlet, $F_3$.  Only the third family can have a tree
level Yukawa coupling to the Higgs (which is also a $D_4$ singlet).

\begin{figure}
  \includegraphics[height=.06\textheight]{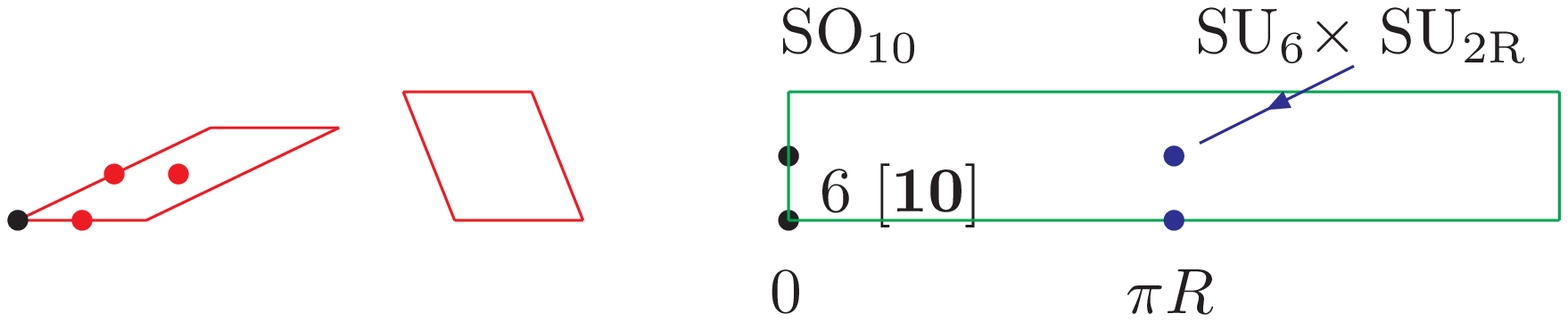}
  \caption{Pure $Z_2$ fixed points with 3 {\bf 10}s residing at each of the $SO_{10}$ invariant fixed points.}
  \label{f:z2fixedpts}
\end{figure}

There is one more twisted sector of the string with fixed points given in Fig. \ref{f:z2fixedpts}.  These are
pure $Z_2$ fixed points with 3 {\bf 10}s residing at each of the $SO_{10}$ invariant fixed points.  This
introduces extra [{\em unwanted}] color triplet states; potential problems for proton decay.
\subsection{Summary of this three family model}

\begin{itemize}
\item Since the third family and the Higgs are derived from the $E_6$ gauge and {\bf 27} hypermultiplet in the 5D
bulk, they have a tree level Yukawa interaction given by
\begin{equation}
\frac{g_5}{\sqrt{\pi R}} \int_0^{\pi R} dy {\bf \overline{27}} \ \Sigma \ {\bf 27} = g \ {\cal H} \ F_3^c \ F_3
\end{equation}
where $g_5$ ($g$) is the 5D (4D) $E_6$ gauge coupling constant evaluated at the compactification scale.  Note,
Pati-Salam symmetry guarantees Yukawa coupling unification for the third generation with
\begin{equation}
g \left[ \approx \sqrt{4 \pi \alpha_G} \approx 0.7 \right] = \lambda_t = \lambda_b = \lambda_\tau =
\lambda_{\nu_\tau} .
\end{equation}

\item The first two families reside at the $Z_2$ fixed points, resulting in a $D_4$ family symmetry.

\item  The massless sector from the 5D bulk also contains the states $ 2( \chi^c + \bar \chi^c) + 3 C $ with $C =
(6, 1, 1)$ under Pati-Salam or a color triplet plus anti-triplet, $(T + \bar T)$, under $SU_{3C}$.  These states
are good news, since they are the necessary Higgs needed for breaking Pati-Salam to the Standard Model gauge
group.   In fact, we have evaluated the 4D effective superpotential (consistent with stringy selection rules) up
to a certain power in Standard Model singlets and we find
\begin{equation}
W \supset S \ C_1 \ C_4 + C_4 \ ( \chi_1^c \ \chi_1^c + \overline{\chi_1^c} \ \overline{\chi_1^c}) + S' \
\overline{\chi_2^c} \ \chi_2^c.
\end{equation}
This is just what is necessary to obtain F and D flat directions breaking $SU_{4C} \times SU_{2L} \times SU_{2R}$
to the $SU_{3C} \times SU_{2L} \times U_{1Y}$ and giving mass to all unwanted charged states under the Standard
Model.

\item  {\em Gauge coupling unification}
In a 4D GUT we have the RG equations given by
\begin{equation}
\frac{2\pi}{\alpha_i(\mu)} \simeq \frac{2\pi}{\alpha^{}_{\rm GUT}}+ b^{\rm MSSM}_i\log\frac{M_{\rm GUT}}{\mu} + 6
\; \delta_{i 3}, \label{eq:4drg}
\end{equation}
where $M_{\rm GUT} \simeq 3 \times 10^{16}$ GeV,  $\alpha_{\rm GUT}^{-1} \simeq 24$ and we have included a
threshold correction at $M_{\rm GUT}$, required in order to fit the low energy data when using the RG equations
from $M_G$ to $M_Z$ at two loop order.  On the other hand, in our effective 5D orbifold GUT we have the following
RG equations:
\begin{eqnarray}
\frac{2\pi}{\alpha_i(\mu)}&\simeq \frac{2\pi}{\alpha_{\rm string}} +b^{\rm MSSM}_i\log\frac{M_{\rm PS}}{\mu} & \nonumber \\
& +(b^{\rm PS}_{++}+b_{\rm brane})_i\log\frac{M_{\rm string}}{M_{\rm PS}} & \nonumber \\  & -\frac{1}{2}(b^{\rm
PS}_{++}+b^{\rm PS}_{--})_i\log\frac{M_{\rm string}}{M_c} \nonumber & \\ & +b^{{\rm E}_6}\left(\frac{M_{\rm
string}}{M_c} -1\right),& \label{eq:5drg}
\end{eqnarray}
where $M_{\rm PS}$ is the PS breaking scale,  $M_c \sim 1/\pi R$ is the 5D compactification scale, $M_{\rm
string}$ is the string scale and $\alpha_{\rm string}$ is the gauge coupling at the string scale.  The
coefficients $b^{\rm PS}_{++}, \ b^{\rm PS}_{--}$ are determined by the massive KK modes with the given parities,
$b^{\rm MSSM}$ is given by the massless modes in the MSSM, and the term $b^{\rm PS}_{++}+b_{\rm brane}$ is
determined by the massless KK modes or the ``brane" (massless fixed point states) which get mass when the
Pati-Salam symmetry is spontaneously broken.
We can use the 4D RG equations to analytically fix some of the parameters in the 5D equations.   Note that the
(++) and ($--$) KK modes are in the overlap of $SO_{10}$ and $SU_6 \times SU_{2R}$, hence $(b^{\rm PS}_{++})_3 =
(b^{\rm PS}_{++})_2$, $(b^{\rm PS}_{--})_3 = (b^{\rm PS}_{--})_2$ and $(b_{\rm brane})_3 = (b_{\rm brane})_2$. As
a consequence, $\alpha_3^{-1} = \alpha_2^{-1}$ and we find $M_{\rm PS} = e^{-3/2} M_G \approx 7 \times 10^{15}$
GeV.
In addition, in the weakly coupled heterotic string we have the boundary condition \begin{equation}  \frac{2
\pi}{\alpha_{\rm string}} = \frac{\pi}{4} \left(\frac{M_{\rm Pl}}{M_{\rm string}}\right)^2 + \frac{1}{2} \
\Delta^{\rm univ}, \label{eq:hetbc} \end{equation} where the first term is the tree level result and the second
term (a universal one loop stringy correction) is negligible.  By assuming the complete freedom to put
non-trivial PS multiplets at either the string or the PS breaking scale we obtain a maximum value of $M_{\rm
string}$ given by  \begin{equation} (M_{\rm string})_{MAX} = e^2 M_G \approx 2 \times 10^{17} \; {\rm GeV}
\end{equation} which (using Eqn. \ref{eq:hetbc}) gives  $\alpha_{\rm string}^{-1} \approx 450$.  This is a
problem.   One possible solution is to discard the weakly coupled heterotic string boundary conditions by
approaching the strong coupling limit.  In the strongly coupled limit we have the Ho\v{r}ava-Witten boundary
conditions given by~\cite{hw}
\begin{equation}
 \frac{2 \pi}{\alpha_{\rm string}} = \frac{1}{2 (4 \pi)^{5/3}M_*\rho} \ \left(\frac{M_{\rm Pl}}{M_*}\right)^2,
\label{eq:witten} \end{equation} where $M_*$ is given in terms of the 11D Newton's constant by $\kappa^{2/3} =
M_*^{-3}$ and $\rho$ is the size of the eleventh dimension.  Now using eq.~\ref{eq:witten}, we find solutions for
$M_{\rm string} \simeq M_* = 2 M_{\rm GUT}, \; M_c \simeq M_{\rm PS} \simeq {\rm e}^{-3/2} M_{\rm GUT}$ with
$n'_{{\bf 2}_R} = n'_{\bf 6+\overline{6}} = 4$ and $M_* \rho \simeq 2$.  Thus we can accommodate gauge coupling
unification.
\end{itemize}

To summarize, the successes of the model are
\begin{enumerate}
\item Successful gauge coupling unification~\cite{unif}

\item An enhanced proton decay rate due to dimension-6 operators with the dominant decay mode $p \rightarrow e^+
\pi^0$. The decay rate for dimension-6 operators is given by~\cite{pdecay}  \begin{equation} \tau (p \rightarrow
e^+ \pi^0) \simeq 3 \times 10^{33} \; \left(\frac{0.015 \ {\rm GeV}^3}{\beta_{\rm lattice}}\right)^2 \;\; {\rm
yrs},
\end{equation} where $\beta_{\rm lattice}$ is an input from lattice calculations of the three quark matrix
element~\cite{lattice}.  Recent results give a range of central values $\beta_{\rm lattice} = 0.007 - 0.015$.
Note, the present experimental bound for this decay mode from Super-Kamiokande is $5.4 \times 10^{33}$ years at
$90\%$ confidence levels~\cite{superk}. Thus this prediction is not yet excluded by the data, but it should be
observed soon.

\item 3rd generation and Higgs in the bulk implies gauge-Yukawa unification (which is a phenomenologically
acceptable relation~\cite{yukawaunif}), and

\item $D_4$ family symmetry, which can generate a family hierarchy and suppress flavor violating interactions.
\end{enumerate}
\begin{eqnarray} (f_1\,f_2\,f_3) \ h_1 \times & & \\ & \left( {\footnotesize
\begin{array}{ccc}
  {\cal O}_2 \tilde S_{\rm e} + S_{\rm e}  &
  {\cal O}_2 \tilde S_{\rm o} + S_{\rm o}  &
 {\cal O}_1 {\cal O}_2 \phi_{\rm e} +  \tilde \phi_{e} \\
  {\cal O}_2 \tilde S_{\rm o} + S_{\rm o}  &
 {\cal O}_2 \tilde S_{\rm e} + S_{\rm e}  &
 {\cal O}_1 {\cal O}_2 \phi_{\rm o} +  \tilde \phi_{\rm o} \\
 \phi'_{e} & \phi'_{\rm o} &   1
\end{array}} \right)\left(
\begin{array}{c}
f_1^c\\
f_2^c\\
f_3^c
\end{array}\right) & \nonumber, \label{yukmat}
\end{eqnarray}
where \begin{equation} S, \ \tilde S, \ \phi, \ \tilde \phi, \ \phi' \end{equation} are singlet operators under
$SO_{10}$, but with non-trivial transformation properties under $D_4$.  The subscripts $\{ e, \ o \}$ correspond
to products of (even, odd) number of $D_4$ doublets.  Also we have defined the two composite operators
\begin{equation} {\cal O}_1 =\overline\chi_1^c\chi^c_\alpha,\qquad {\cal O}_2 =\overline\chi_2^c\chi^c_\alpha,
\end{equation} where the group indices are arranged in all possible ways.  Hence, the structure of the Yukawa
matrix is determined by the $D_4$ family symmetry and the PS symmetry breaking VEVs.

\subsection{So What's the Problem?}

There are two potential problems and, unfortunately, there is one killer problem.
\begin{itemize}
\item  We have a number of exotic states with fractional electric charge.  \item We have 6 extra multiplets in
complete {\bf 10} dimensional representations of $SO_{10}$.  \item  R parity is broken!
\end{itemize}
Perhaps there is a solution in a different $Z_N \times Z_2$ orbifold construction.

\subsubsection{Exotics}

The good news is that there are very few exotic states AND they come in vector-like pairs.  We list the exotics
with their PS quantum numbers below.
\begin{eqnarray}
q_i = (4, 1, 1), & \bar q_i = (\bar 4, 1, 1); & i = 1,2 \\
D_j^l = (1, 2, 1), &  D_j^r = ( 1, 1, 2); &  j = 1, \cdots, 4 . \nonumber
\end{eqnarray}
Hence, in principle, all these states with fractional electric charge could get mass near the string scale, once
singlet scalars obtain non-vanishing VEVs.  The potential problem with these states is that NO fractional charged
particles have ever been seen.   They have not been produced in any high energy accelerators, thus they must be
heavier than about several hundred GeV.   On the other hand, they have never been seen in mass spectrometer
searches of thousands of gallons of sea water, looking for exotic states of hydrogen.  The limits on exotic
charged matter are of order $10^{-24}$ in abundance compared with normal hydrogen~\cite{smith}.   Of course, by
charge conservation the lightest fractional charged state, if it existed, would be stable.  Thus such states
could not have been produced in the early universe.  Hence they should be heavier than any reheat temperature
after inflation.  In this case, their present abundance would be low enough, so that they would not be observable
today.

The problem for this particular model is that one of the exotics is an oscillator mode, while its vector-like
partner is not. We have searched for effective mass operators, to order power 8 in $SO_{10}$ singlets and we
cannot find an operator which would give mass to the oscillator modes.   I should note, that in other models
discussed in the paper~\cite{Kobayashi:2004ya},  all of the exotics were non-oscillator modes and all could in
principle obtain mass from effective mass operators.   Thus we conclude that this is a problem specific to this
particular model and not generic.   However, for this model this problem is a killer!

\subsubsection{R parity}

The superpotential contains terms of the form
\begin{eqnarray}
W \supset & S \ C_1 \ C_4 + C_1 \ (\overline{\chi_1^c} \ \overline{\chi_2^c}) &  \\
&  C_4 \ (F_3^c \ \chi_\alpha^c + (F_A^c \ F_B^c +  F_A \ F_B)) + S' \ (\overline{\chi^c} \ \chi^c) .&
\label{rparity}  \nonumber
\end{eqnarray}

In the MSSM, R parity is equivalent to family reflection symmetry [FRS] in which all ordinary family superfields
are odd ($F \rightarrow - F$) and the Higgs superfield is even (${\cal H} \rightarrow {\cal H}$).   In order to
preserve this symmetry, even after Pati-Salam symmetry is broken, we need that the Higgs which break PS are even
under FRS, i.e.  ($\chi^c \rightarrow \chi^c$).   Clearly with the superpotential (Eqn. \ref{rparity}) this
symmetry is manifestly broken.  Hence R parity is broken in the effective low energy field theory.  This may or
not be a problem.  It depends on whether or not the dimension 4, R parity violating operators can be sufficiently
suppressed to withstand the known limits.

However, we suggest that it may be possible to find models which have an exact low energy R parity.   In this
model cubic terms in the superpotential (Eqn. \ref{rparity}) of the form $T_4 ( U_3 \ T_2 +  T_1 \ T_1)$, where
$U_i, \ i = 1,2,3$ refers to untwisted sector states and $T_i, \ i = 1,2,3,4,5$ refers to $Z_6$ twisted sector
states, are allowed by the stringy selection rules. The fields $F_3^c$ and $F_A, \ F^c_A$ come from the $U_3$,
$T_1$ sectors, respectively. Perhaps in a different $Z_N \times Z_2$ orbifold we would have the first term, but
the second term would be forbidden at cubic order. In this case,  the superpotential (Eqn. \ref{rparity}) might
be replaced by
\begin{eqnarray}
W \supset & S \ C_1 \ C_4 + C_1 \ (\overline{\chi_1^c} \ \overline{\chi_2^c}) &  \\
&  C_4 \ (F_3^c \ \chi_\alpha^c + \tilde S \ (F_A^c \ F_B^c +  F_A \ F_B)) + S' \ (\overline{\chi^c} \ \chi^c). &
\label{rparity2}  \nonumber
\end{eqnarray}
with $\tilde S$ odd under FRS.  Then, as long as $\tilde S$ does not develop a vacuum expectation value, R parity
would be an exact symmetry of the effective low energy field theory.

\subsection{Conclusions}

\begin{itemize}

\item We have presented a new 3 family $E_6$ orbifold GUT which has an ultra-violet completion in the $E_8 \times
E_8$ heterotic string.

\item  The Yukawa coupling, of the 3rd generation of quarks and leptons to the Higgs, satisfies gauge-Yukawa
unification.   This relation is very predictive.   Moreover it is not ruled out.

\item The model has a $D_4$  family symmetry.  This symmetry can naturally explain the family hierarchy and also
suppress flavor violation

\item  The model includes a small number of vector-like exotics.   Unfortunately, in this particular model, it is
not possible to give all the exotics mass.   This is the super killer problem for this construction!   The good
news is that this problem was very specific to this particular model.  Other models we have constructed do not
have this problem.

\item  The model contains extra color triplets.   These are a nuisance;  they are not necessarily a problem. In
this model, these additional color triplets mediate R parity violating processes.

 \item  The model has NO R parity.   This is aesthetically unappealing.  It however may be experimentally
 acceptable.
\end{itemize}

Finally, we suggest that possible solutions to the above problems may be found in alternative $Z_N \times Z_2$
orbifold constructions.   We are continuing to look in this direction.

\begin{theacknowledgments}
I would like to thank the organizers of PASCOS 2005 for hosting such a wonderful conference and for giving me the
opportunity to present this talk.
\end{theacknowledgments}

\end{document}